# FWM in One-dimensional Nonlinear Photonic Crystal and Theoretical Investigation of Parametric Down Conversion Efficiency (Steady State Analysis)

**M. Boozarjmehr[1] and A. Rostami [2]**

1. Photonics Group, Physics Faculty, University of Tabriz, Tabriz 51664,Iran
2. Photonics and Nanocrystals Research Lab., (PNRL), Faculty of Electrical and Computer Engineering, University of Tabriz, Tabriz 51664, Iran

**Abstract-** We study the light propagation in one-dimensional photonic crystal via nonlinear Four Wave Mixing (FWM) process. The linear and nonlinear refractive indexes are approximated with the first Fourier harmonic term. A system of the nonlinear coupled mode equations (NLCMEs), including pump fields depletion is derived for FWM process and steady state analysis is presented numerically. Also some of important system parameters' effects on FWM process are investigated. It has been shown that, although we have considered pump depletion and used a $\chi^{(3)}$ medium which is small compared to $\chi^{(2)}$, conversion efficiency is enhanced at least 100 times to the previous works which had used undepleted pump approximation and $\chi^{(2)}$ medium.

**I. Introduction-** Parametric interactions in nonlinear periodic structures play an important role in all-optical networks. Four Wave Mixing process is one of these interactions which has long been studied [1-3]. Recently, FWM process is used in interesting areas of quantum information technology to generate single and entangled photons via nonlinear photonic crystals [4-6].

In most of these crystals polarization of photons is used to generate entangled photon pairs via parametric interactions such as FWM.

These crystals include BBO and GaAs bulk materials, but since the nonlinear susceptibility of GaAs is much greater than BBO crystals, which affect the conversion efficiency of FWM process, GaAs crystals are widely used in quantum optics areas of research, but because of lack of birefringence, phase matching condition is not satisfied easily, so in order to obtain phase matching condition easily, nonlinear photonic crystals are used instead of bulk GaAs.

FWM is really a photon-photon scattering process, during which two photons from a relatively high-intensity beam, called pump beams scatter through third order $\chi^{(3)}$ nonlinearity of a material to generate two correlated photons called signal and idler photons respectively [3, 7, 8]. In homogeneous nonlinear media (such as bulk material), efficient exchange of energy between interacting modes of the electromagnetic field is determined by the linear and nonlinear susceptibilities of the medium. So, successful achievement of the proposed applications strongly depends on the nonlinearity strength and the medium structure. But these materials suffer from several problems which some of them are mentioned below:

1. $\chi^{(3)}$ nonlinearity is usually small compared to $\chi^{(2)}$.
2. Wavelengths of signal and idler photons are close to pump wavelengths.
3. In the case of small conversion efficiency, even small amounts of pump beam scattering generates large background count rates that mask the detection of correlations between signal and idler photons. (on the other hand scattering of the pump fields tends to mask the desired quantum effects).

1.Author to whom any comment should be addressed : maryamboozarjmehr@gmail.com

Many of the problems associated with $\chi^{(3)}$ nonlinear optical materials can be eliminated by using suitable structures such as single mode optical fibers. These optical fibers have extremely low loss, small confinement cross section and can be as long as several kilometers. The nonlinearity is an off-resonance $\chi^{(3)}$ Kerr effect with an ultra fast frequency response extending from dc to well above 10 THz. Although weak, it can give rise to very large nonlinear effect in long fibers.
It is obvious that the material's permittivity determines how phase matched is a given parametric process, whereas the actual coupling of energy between the modes is a function of the material's nonlinear polarizability. In an attempt to circumvent material constraints (second alternative), much works have been focused on the possibility of using periodic media to mediate nonlinear processes. Some of basic important works proposed the introduction of periodic structure into the linear and nonlinear material properties to aid in phase matching parametric interactions [6, 9-12].
The introduction of the periodic nonlinear modulation leads to both flexibility in phase matching and also makes accessible a material's largest nonlinear coefficient. It has been shown that periodic modulation of a nonlinear material's refractive index can lead to enhanced conversion efficiencies in parametric processes.
Photonic crystals were first conceived by John and Yablonovitch [13, 14], and have been widely used in all fields of optics, so especial arrangement of linear and nonlinear index of refraction can help us to modify the conversion efficiency of FWM process.
In this paper, we propose a complete set of coupled wave equations describing FWM in one-dimensional nonlinear photonic crystal for the first time. The derived relations include all system parameters and input status. Our consideration concentrate on $1.55\mu m$ which is interesting for optical communication. Also, the proposed periodic structure can be imagined as nonlinear fiber Bragg Grating. After derivation of the coupled wave equations for all field components (forward and backward components), numerical methods have been used to simulate of the process. Simulation results for conversion efficiency are presented both for co-propagating and counter-propagating signal fields. Also, we try to enhance conversion efficiencies using FWM process.
This paper is organized as follows:
In section (II) mathematical model and derivation of FWM process in one dimensional photonic crystal is discussed. Section (III) includes numerical evaluation of equations and some of important system parameters' effects on FWM process are investigated. Finally this paper ends with summary of the work and conclusion in section (IV).

**II. Mathematical Model-** We now present a mathematical model for FWM process in 1D photonic crystal. Schematic sketch of a nonlinear photonic crystal for modeling FWM process is illustrated in Fig(1).

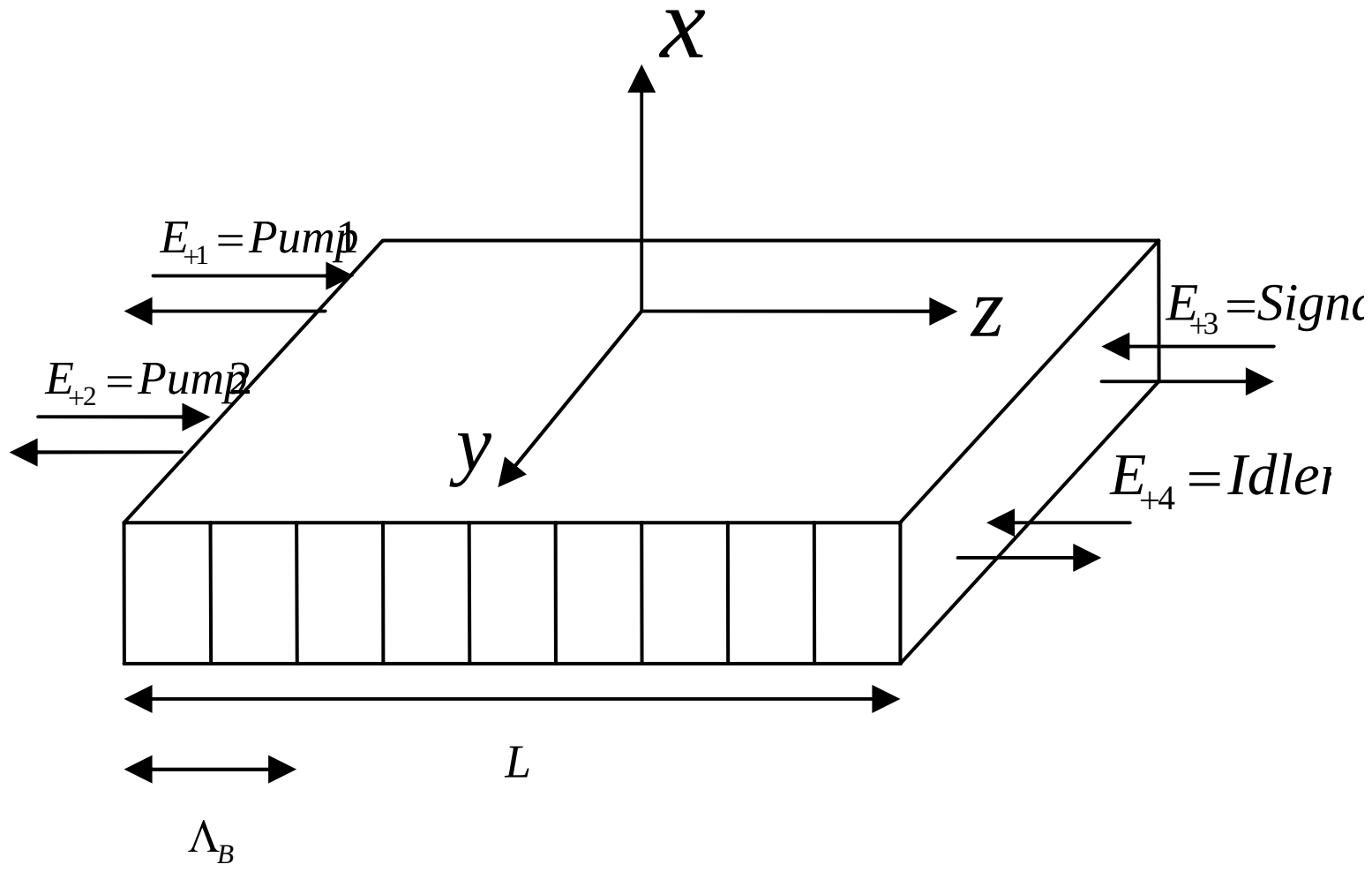

Fig (1): One-dimensional nonlinear photonic crystal and the indexes of refraction distribution

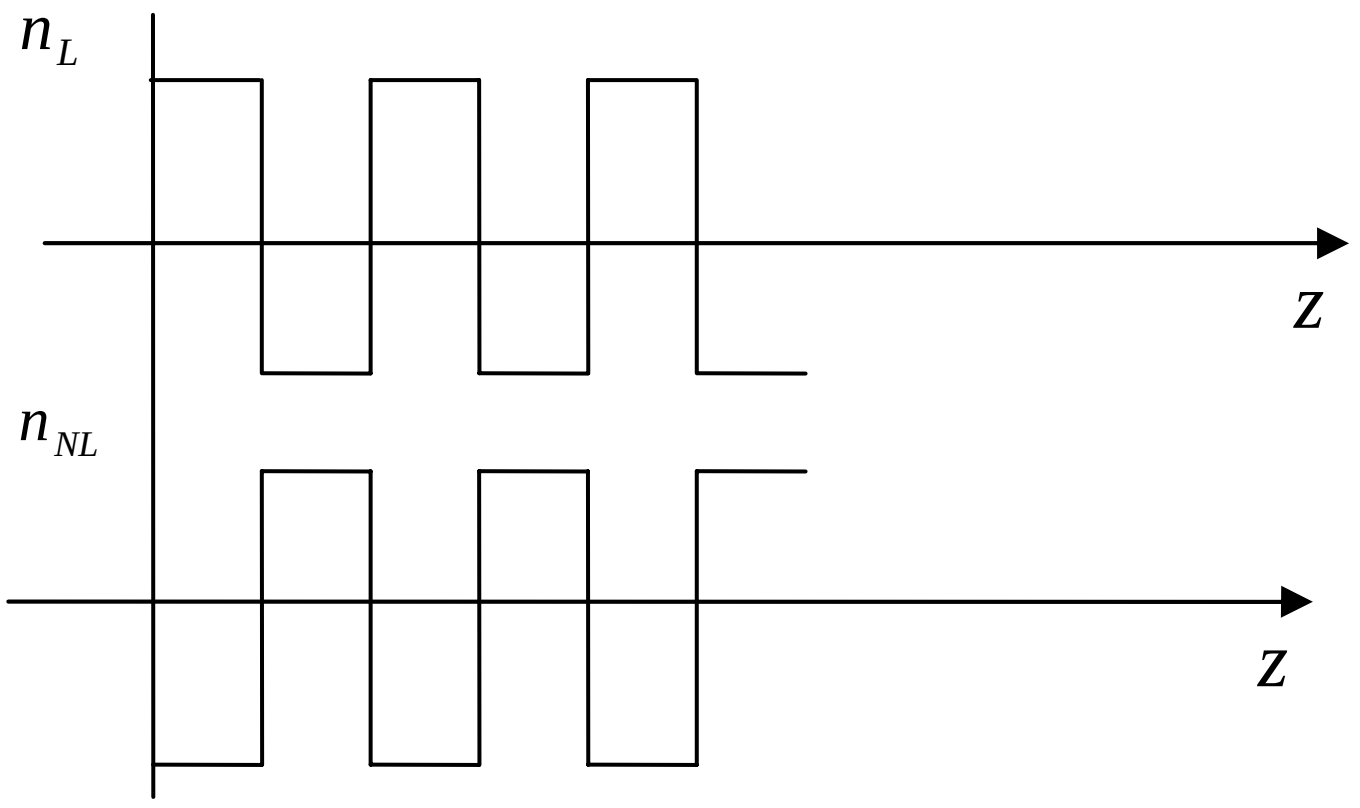

Fig (2): Typical profiles of linear and nonlinear refractive indexes

For the proposed structure the refractive index profile is given as follows:

$$n = n_0 + a_1 \cos(2k_0 + \delta)z - a_2 |E|^2 \cos(2k_0 + \delta)z, \qquad (1)$$

where $n_0$, $a_1$, $k_0$, $\delta$, $a_2$ and $E$ are the average refractive index of crystal, the first harmonic coefficient of Fourier expansion for the linear index of refraction, average incident wave vector, phase mismatch between incident wave vector and periodic structure's wave vector, the first harmonic coefficient of Fourier expansion for the nonlinear index of refraction and the applied electric field, respectively. Re. (1) stands that both linear and nonlinear components of the refractive index are chosen periodic as shown in Fig (2).
We have written the following field distribution for FWM process in the periodic structure,

$$\begin{aligned} E = {} & (E_{+1} e^{ik_1 z} + E_{-1} e^{-ik_1 z})e^{-i\omega_1 t} + (E_{+2} e^{ik_2 z} + E_{-2} e^{-ik_2 z})e^{-i\omega_2 t} + \\ & (E_{+3} e^{-ik_3 z} + E_{-3} e^{ik_3 z})e^{-i\omega_3 t} + (E_{+4} e^{-ik_4 z} + E_{-4} e^{ik_4 z})e^{-i\omega_4 t} + c.c., \end{aligned} \qquad (2)$$

where $E_{\pm i}$, $k_i$, and $\omega_i$ are amplitudes of the forward and backward pump , signal and idler fields, their wave vectors and frequencies for all components, respectively. In writing the electric field distribution, phase mismatch condition between four wave vectors should be satisfied,

$$\Delta k = k_1 + k_2 - k_3 - k_4 \qquad (3)$$

which $k_1$ , $k_2$ , $k_3$ and $k_4$ are two pump, signal and idler wave vectors respectively.

Nonlinear polarization is [6],

$$P_{NL} = \varepsilon_0 [A(z)(E_i . E_j{}^*) E_k + \frac{1}{2} B(z)(E_i . E_j) E_k{}^*] + c.c.,$$
(4)

where $A(z)$ and $B(z)$ are nonlinear related to nonlinear medium distribution profile and given as follows,

$$A(z) = B(z) = -A[e^{i(2k_0+\delta)z} + e^{-i(2k_0+\delta)z}], \tag{5}$$

where

$$k_0 = \frac{k_1 + k_2 + k_3 + k_4}{4}. \tag{6}$$

Now, for obtaining the coupled wave equations, the electric filed and the nonlinear polarization should satisfy the Maxwell's wave equation,

$$\frac{\partial^2 E}{\partial z^2} - \frac{n^2}{c^2}\frac{\partial^2 E}{\partial t^2} = \mu_0 \frac{\partial^2 P_{NL}}{\partial t^2}, \tag{7}$$

where $c$ is speed of light in vacuum and $n$ is the refractive index of medium.
Because of small perturbation in the refractive index, the following approximation is used for refractive index in Maxwell's wave equation,

$$n^2 = n_0{}^2 + n_0 a_1 (e^{i(2k_0+\delta)z} + e^{-i(2k_0+\delta)z}) - n_0 a_2 (e^{i(2k_0+\delta)z} + e^{-i(2k_0+\delta)z})|E|^2 \tag{8}$$

Finally, after substitution (2), (4) and (8) in Eq. (7) ,using slowly varying function approximation and doing some mathematical simplifications, the following coupled wave equations are obtained,

$$\frac{\partial E_{+1}}{\partial z} = \frac{i\omega_1^2 n_0 a_1 E_{-1} e^{i\delta z}}{2k_1 c^2} - \frac{i\omega_1^2 (n_0 a_2 + \frac{3}{4}A)}{k_1 c^2} \times \tag{9}$$
$$\{\alpha_1 E_{-1} e^{i\delta z} + \alpha_2 E_{+1} e^{i\delta z} + \alpha_3 E_{+1} e^{-i\delta z} + 2E_{-3}^* E_{-4}^* E_{+2} e^{i\delta z} e^{i\Delta kz}\},$$

where

$$\alpha_1 = 2|E_{+1}|^2 + 2|E_{+2}|^2 + 2|E_{+3}|^2 + 2|E_{+4}|^2 + |E_{-1}|^2 + 2|E_{-2}|^2 + 2|E_{-3}|^2 + 2|E_{-4}|^2,$$
$$\alpha_2 = 2E_{+3}E_{-3}^* + 2E_{+4}E_{-4}^* + 2E_{-2}E_{+2}^*, \tag{10}$$
$$\alpha_3 = E_{+1}E_{-1}^* + 2E_{+2}E_{-2}^* + 2E_{-3}E_{+3}^* + 2E_{-4}E_{+4}^*.$$

Eq. (9) illustrates the coupled wave equation for the first pump field propagating from left to in crystal

$$-\frac{\partial E_{-1}}{\partial z} = \frac{i\omega_1^2 n_0 a_1 E_{+1} e^{-i\delta z}}{2k_1 c^2} - \frac{i\omega_1^2 (n_0 a_2 + \frac{3}{4}A)}{k_1 c^2} \times \tag{11}$$
$$\{\Gamma_1 E_{+1} e^{-i\delta z} + \Gamma_2 E_{-1} e^{-i\delta z} + \Gamma_3 E_{-1} e^{i\delta z} + 2E_{-3}E_{-4}E_{+2}^* e^{-i\delta z} e^{-i\Delta kz}\},$$

where

$$\Gamma_1 = |E_{+1}|^2 + 2|E_{+2}|^2 + 2|E_{+3}|^2 + 2|E_{+4}|^2 + 2|E_{-1}|^2 + 2|E_{-2}|^2 + 2|E_{-3}|^2 + 2|E_{-4}|^2,$$
$$\Gamma_2 = 2E_{+2}E_{-2}^* + 2E_{-3}E_{+3}^* + 2E_{-4}E_{+4}^*, \tag{12}$$
$$\Gamma_3 = E_{-1}E_{+1}^* + 2E_{-2}E_{+2}^* + 2E_{+3}E_{-3}^* + 2E_{+4}E_{-4}^*.$$

Eq. (11) illustrates the coupled wave equation for the backward component of the first pump field propagating from right to left in crystal.

$$\frac{\partial E_{+2}}{\partial z} = \frac{i\omega_2^2 n_0 a_1 E_{-2} e^{i\delta z}}{2k_2 c^2} - \frac{i\omega_2^2 (n_0 a_2 + \frac{3}{4}A)}{k_2 c^2} \times \tag{13}$$
$$\{\beta_1 E_{-2} e^{i\delta z} + \beta_2 E_{+2} e^{i\delta z} + \beta_3 E_{+3} e^{-i\delta z} + 2E_{-3}^* E_{-4}^* E_{+1} e^{i\delta z} e^{i\Delta kz}\},$$

where

$$\beta_1 = 2|E_{+1}|^2 + 2|E_{+2}|^2 + 2|E_{+3}|^2 + 2|E_{+4}|^2 + 2|E_{-1}|^2 + |E_{-2}|^2 + 2|E_{-3}|^2 + 2|E_{-4}|^2,$$
$$\beta_2 = 2E_{+3}E_{-3}^* + 2E_{+4}E_{-4}^* + 2E_{-1}E_{+1}^*, \tag{14}$$
$$\beta_3 = 2E_{+1}E_{-1}^* + E_{+2}E_{-2}^* + 2E_{-3}E_{+3}^* + 2E_{-4}E_{+4}^*.$$

Eq. (13) illustrates the coupled wave equation for the second pump field propagating from left to right in crystal.

$$-\frac{\partial E_{-2}}{\partial z} = \frac{i\omega_2^2 n_0 a_1 E_{+2} e^{-i\delta z}}{2k_2 c^2} - \frac{i\omega_2^2 (n_0 a_2 + \frac{3}{4}A)}{k_2 c^2} \times \tag{15}$$
$$\{\theta_1 E_{+2} e^{-i\delta z} + \theta_2 E_{-2} e^{-i\delta z} + \theta_3 E_{-3} e^{i\delta z} + 2E_{-3} E_{-4} E_{+1}^* e^{-i\delta z} e^{-i\Delta kz}\},$$

where

$$\theta_1 = 2|E_{+1}|^2 + |E_{+2}|^2 + 2|E_{+3}|^2 + 2|E_{+4}|^2 + 2|E_{-1}|^2 + 2|E_{-2}|^2 + 2|E_{-3}|^2 + 2|E_{-4}|^2,$$
$$\theta_2 = 2E_{+2}E_{-2}^* + 2E_{-3}E_{+3}^* + 2E_{-4}E_{+4}^*, \tag{16}$$
$$\theta_3 = 2E_{-1}E_{+1}^* + E_{-2}E_{+2}^* + 2E_{+3}E_{-3}^* + 2E_{+4}E_{-4}^*.$$

Eq. (15) illustrates the coupled wave equation for the backward component of the second pump field propagating from right to left in crystal.

$$-\frac{\partial E_{+3}}{\partial z} = \frac{i\omega_3^2 n_0 a_1 E_{-3} e^{-i\delta z}}{2k_3 c^2} - \frac{i\omega_3^2 (n_0 a_2 + \frac{3}{4}A)}{k_3 c^2} \times \tag{17}$$
$$\{\gamma_1 E_{-3} e^{-i\delta z} + \gamma_2 E_{+3} e^{-i\delta z} + \gamma_3 E_{+3} e^{i\delta z} + 2E_{+1} E_{+2} E_{-4}^* e^{-i\delta z} e^{i\Delta kz}\},$$

where

$$\gamma_1 = 2|E_{+1}|^2 + 2|E_{+2}|^2 + 2|E_{+3}|^2 + 2|E_{+4}|^2 + 2|E_{-1}|^2 + 2|E_{-2}|^2 + |E_{-3}|^2 + 2|E_{-4}|^2,$$
$$\gamma_2 = 2E_{+1}E_{-1}^* + 2E_{+2}E_{-2}^* + 2E_{-4}E_{+4}^*, \tag{18}$$
$$\gamma_3 = 2E_{-1}E_{+1}^* + 2E_{-2}E_{+2}^* + E_{+3}E_{-3}^* + 2E_{+4}E_{-4}^*.$$

Eq. (17) illustrates the coupled wave equation for the signal field propagating from right to left in crystal.

$$\frac{\partial E_{-3}}{\partial z}=\frac{i\omega_3^2 n_0 a_1 E_{+3} e^{i\delta z}}{2k_3 c^2}-\frac{i\omega_3^2 (n_0 a_2+\frac{3}{4}A)}{k_3 c^2}\times \tag{19}$$
$$\{\psi_1 E_{+3} e^{i\delta z}+\psi_2 E_{-3} e^{i\delta z}+\psi_3 E_{-3} e^{-i\delta z}+2E_{-4}E_{+1}^{*}E_{+2}^{*}e^{i\delta z}e^{-i\Delta kz}\},$$

where

$$\begin{aligned}\psi_1 &= 2|E_{+1}|^2+2|E_{+2}|^2+|E_{+3}|^2+2|E_{+4}|^2+2|E_{-1}|^2+2|E_{-2}|^2+2|E_{-3}|^2+2|E_{-4}|^2,\\ \psi_2 &= 2E_{+4}E_{-4}^{*}+2E_{-2}E_{+2}^{*}+2E_{-1}E_{+1}^{*},\\ \psi_3 &= 2E_{+1}E_{-1}^{*}+2E_{+2}E_{-2}^{*}+2E_{-3}E_{+3}^{*}+2E_{-4}E_{+4}^{*}.\end{aligned} \tag{20}$$

Eq. (19) illustrates the coupled wave equation for the backward component of the signal field propagating from left to right in crystal.

$$-\frac{\partial E_{+4}}{\partial z}=\frac{i\omega_4^2 n_0 a_1 E_{-4} e^{-i\delta z}}{2k_4 c^2}-\frac{i\omega_4^2 (n_0 a_2+\frac{3}{4}A)}{k_4 c^2}\times \tag{21}$$
$$\{\eta_1 E_{-4} e^{-i\delta z}+\eta_2 E_{+4} e^{-i\delta z}+\eta_3 E_{+4} e^{i\delta z}+2E_{+1}E_{+2}E_{-3}^{*}e^{-i\delta z}e^{i\Delta kz}\},$$

where

$$\begin{aligned}\eta_1 &= 2|E_{+1}|^2+2|E_{+2}|^2+2|E_{+3}|^2+2|E_{+4}|^2+2|E_{-1}|^2+2|E_{-2}|^2+2|E_{-3}|^2+|E_{-4}|^2,\\ \eta_2 &= 2E_{+1}E_{-1}^{*}+2E_{+2}E_{-2}^{*}+2E_{-3}E_{+3}^{*},\\ \eta_3 &= 2E_{-1}E_{+1}^{*}+2E_{-2}E_{+2}^{*}+2E_{+3}E_{-3}^{*}+E_{+4}E_{-4}^{*}.\end{aligned} \tag{22}$$

Eq. (21) illustrates the coupled wave equation for the idler field propagating from right to left in the medium.

$$\frac{\partial E_{-4}}{\partial z}=\frac{i\omega_4^2 n_0 a_1 E_{+4} e^{i\delta z}}{2k_4 c^2}-\frac{i\omega_4^2 (n_0 a_2+\frac{3}{4}A)}{k_4 c^2}\times \tag{23}$$
$$\{\mathrm{K}_1 E_{+4} e^{i\delta z}+\mathrm{K}_2 E_{-4} e^{i\delta z}+\mathrm{K}_3 E_{-4} e^{-i\delta z}+2E_{-3}E_{+1}^{*}E_{+2}^{*}e^{i\delta z}e^{-i\Delta kz}\},$$

where

$$\begin{aligned}\mathrm{K}_1 &= 2|E_{+1}|^2+2|E_{+2}|^2+2|E_{+3}|^2+|E_{+4}|^2+2|E_{-1}|^2+2|E_{-2}|^2+2|E_{-3}|^2+2|E_{-4}|^2,\\ \mathrm{K}_2 &= 2E_{+3}E_{-3}^{*}+2E_{-2}E_{+2}^{*}+2E_{-1}E_{+1}^{*},\\ \mathrm{K}_3 &= 2E_{+1}E_{-1}^{*}+2E_{+2}E_{-2}^{*}+2E_{-3}E_{+3}^{*}+2E_{-4}E_{+4}^{*}.\end{aligned} \tag{24}$$

Eq. (23) illustrates the coupled wave equation for the backward component of the idler field propagating from left to right in crystal.

We have solved these equations numerically and the effect of system parameters on conversion efficiency (energy transfer efficiency from pump to signal field) is considered and illustrated in the next section. The conversion efficiency for forward traveling signal components can be defined as follows,

$$\eta_{+} = \frac{P_{+3}(0)}{P_{+1}(0)}, \tag{25}$$

We will call this efficiency co-propagation efficiency in the next sections. $P_{+3}(0)$ and $P_{+1}(0)$ stands for the forward pump and signal power respectively, which signal and pump fields are applied from right hand side and left hand side to our system respectively. Also, we define conversion efficiency parameter for the backward signal power as follows,

$$\eta_{-} = \frac{P_{-3}(L)}{P_{+1}(0)}, \tag{26}$$

where $P_{-3}(L)$ is the backward signal power at the right hand side of crystal.

**III. Simulation Results and Discussion-** In this section we present some of numerical results for both co-propagating and counter-propagating efficiencies in FWM process.

Our simulations show that the co-propagating efficiency is larger than the counter-propagating efficiency; this is due to the basic principles of energy transfer between forward and backward propagating modes in Brag gratings.

Fig (3) shows the effect of number of periods on co-propagation conversion efficiency. Conversion efficiency is decreased by decreasing of the nonlinear refractive index coefficient. As it is shown in Fig (3) we have increased the layers f crystal up to 600 layers, in simulation process we observed that by choosing the number of layers to more than 600 layers, co-propagating efficiency has oscillatory behavior, so it is necessary to keep in mind that the conversion efficiency is nonlinearly dependant on the number of periods and this subject should be considered in design process of a photonic crystal to avoid decreasing of the conversion efficiency.

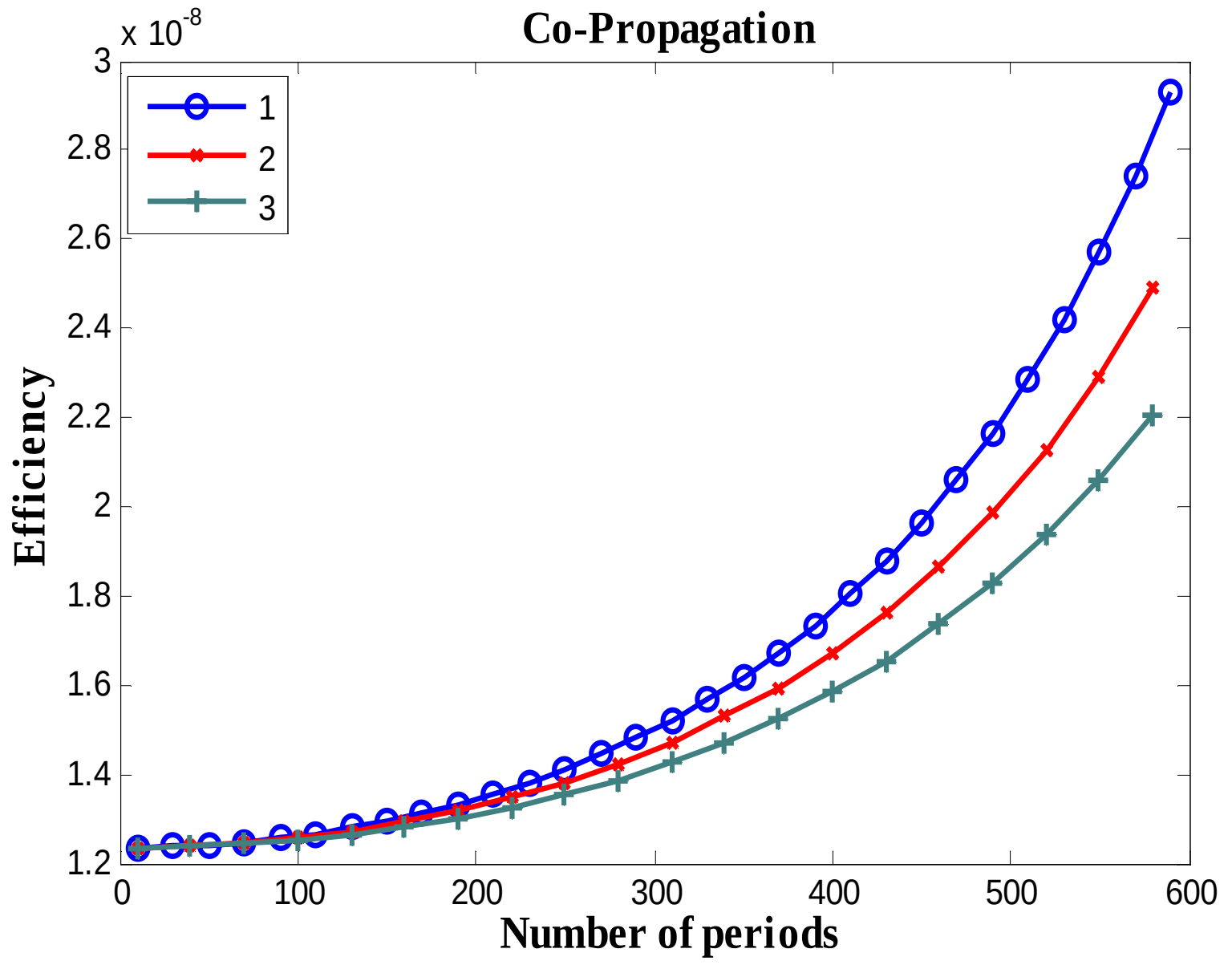

Fig. (3): Efficiency vs. number of periods for different values of $a_2$

1) $a_2 = -10^{-15}$, 2) $a_2 = -2\times10^{-15}$, 3) $a_2 = -3\times10^{-15}$, $n_0 = 3.45$, $\Delta k = 0(m^{-1})$, $a_1 = 0.001$, $\delta = 0(m^{-1})$

Fig (4) illustrates the effect of phase mismatch between wave vectors of four optical fields (two pumps, signal and idler fields) on conversion efficiency. It is shown that by increasing of phase mismatch, the conversion efficiency is decreased, due to the weak energy transfer between propagating modes in this case.

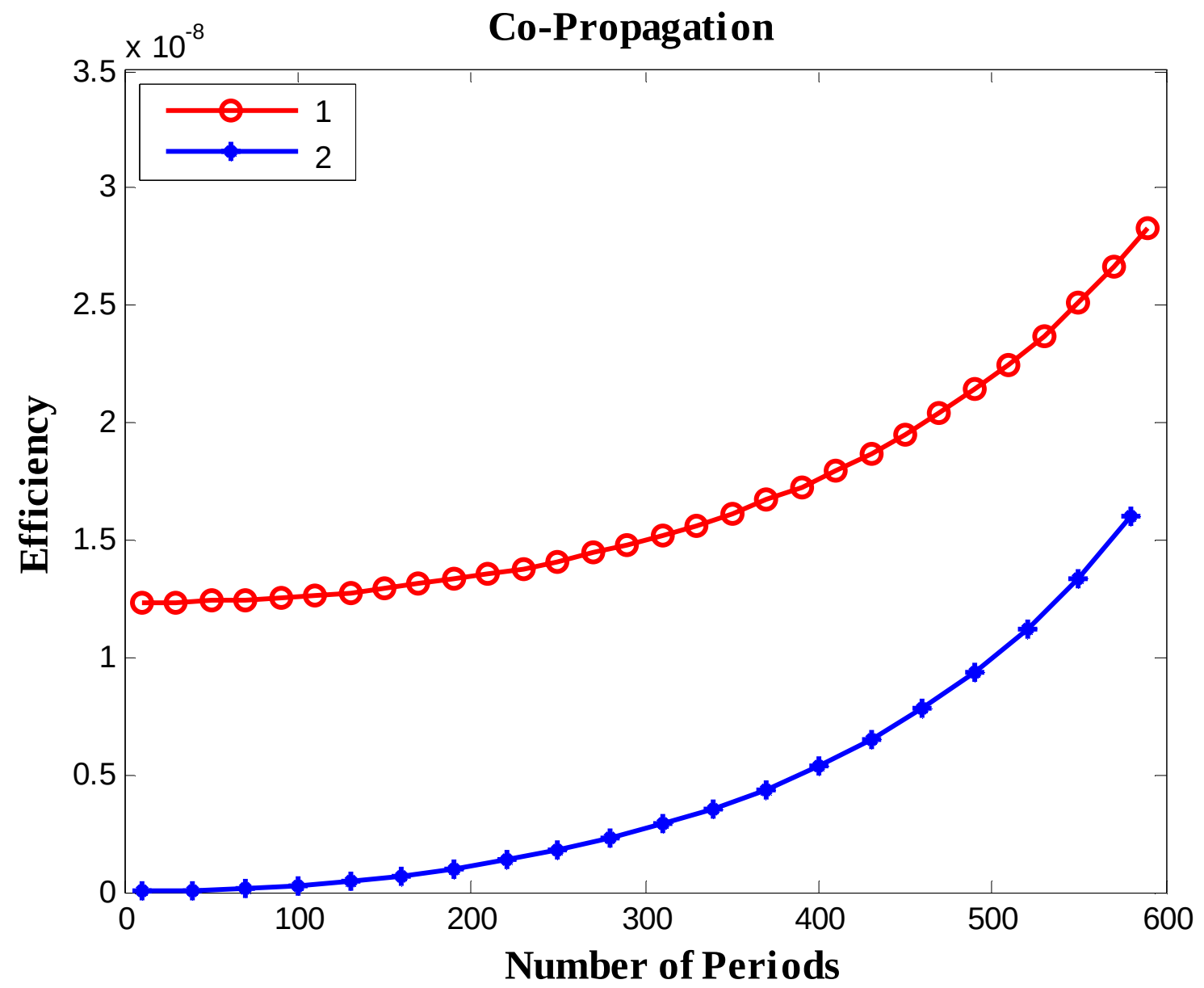

Fig. (4): Efficiency vs. number of periods for different values of $\Delta k$

1) $\Delta k = 0.0001k_0(m^{-1})$, 2) $\Delta k = 0.001k_0(m^{-1})$, 3) $a_2 = -10^{-15}$, $n_0 = 3.45$, $\delta = 0(m^{-1})$, $a_1 = 0.001$

Figs (5, 6) demonstrate the effect of phase mismatch between the average wave vector of optical fields and Grating wave vector on the conversion efficiency, as $\delta$ increases the conversion efficiency decreases. So in order to obtain high efficiencies in Bragg gratings, $\delta$ should be considered as small as possible. Also careful choice of number of layers is important due to the nonlinear dependence between number of layers and the conversion efficiency.

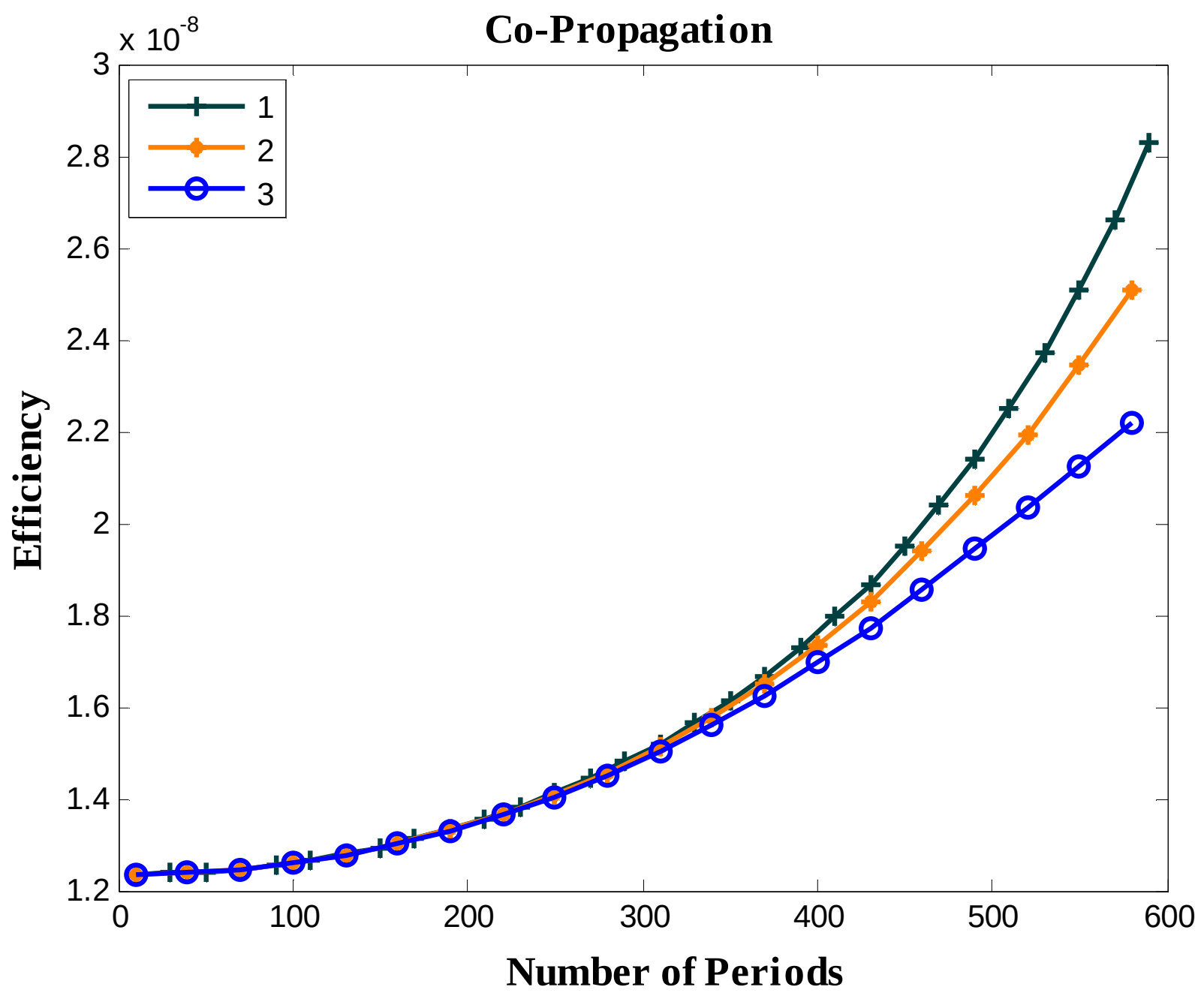

Fig. (5): Efficiency vs. number of periods for different values of $\delta$

1) $\delta = 0.001k_0(m^{-1})$, 2) $\delta = 0.002k_0(m^{-1})$, 3) $\delta = 0.003k_0(m^{-1})$, $n_0 = 3.45$, $\Delta k = 0(m^{-1})$, $a_1 = 0.001$, $a_2 = -10^{-15}$

The effect of between Grating wave vector and the average wave vector of four optical fields on conversion efficiency in broad ranges is shown in Fig. (4).

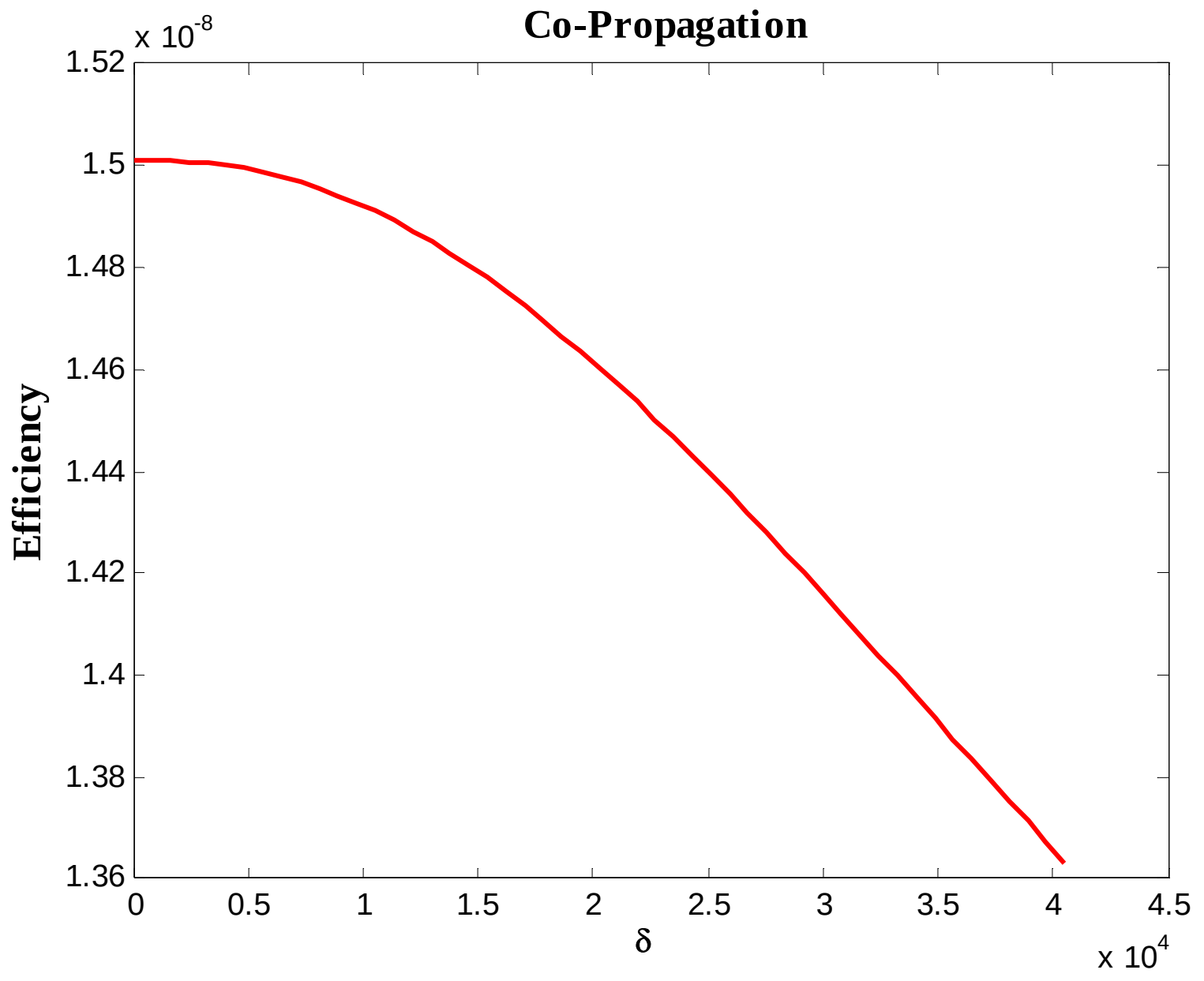

Fig. (6): Efficiency vs. phase mismatch between grating and average applied fields' wave vectors
$n_0 = 3.45, \Delta k = 0(m^{-1}), a_1 = 0.001, a_2 = -10^{-15}, N = 300$

Fig (7)illustrates the effect of the nonlinear refractive index coefficient on the conversion efficiency, as we have defined the nonlinear refractive index's coefficient with a negative sign in relation (1), we have drawn efficiency vs. absolute value of $a_2$ ,taking into account this point, it is shown that by increasing the nonlinear refractive index's coefficient, the conversion efficiency increases.

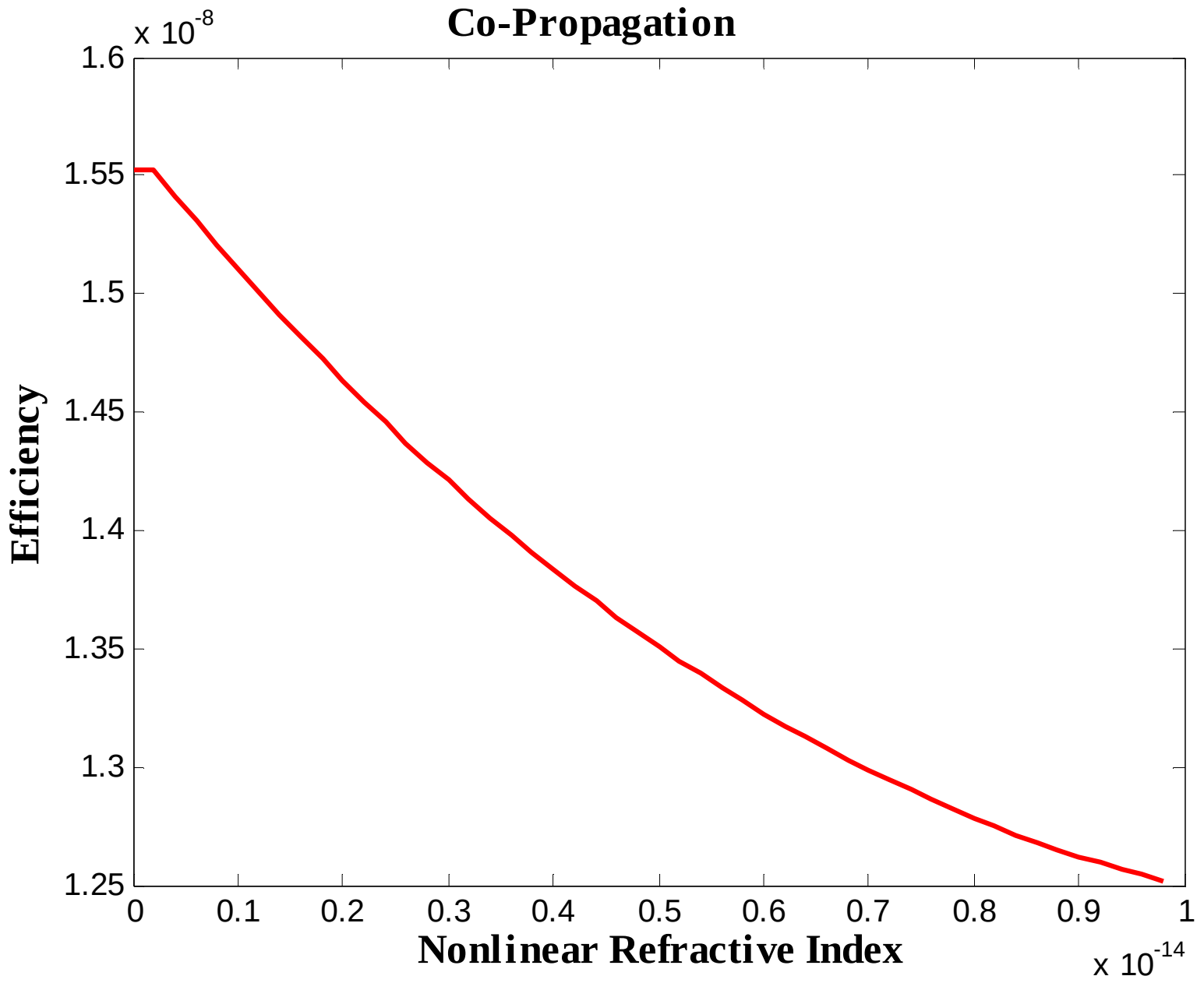

Fig. (7): Efficiency vs. absolute value of nonlinear refractive index's coefficient
$n_0 = 3.45, \Delta k = 0(m^{-1}), a_1 = 0.001, \delta = 0(m^{-1}), N = 300$

The same simulations have been illustrated for counter-propagation case.

Fig (8) and (9) show the effect of phase mismatch between the average wave vectors of the optical fields and grating wave vector on the conversion efficiency, and as it is expected, by increasing of $\delta$,the counter-propagating efficiency decreases.

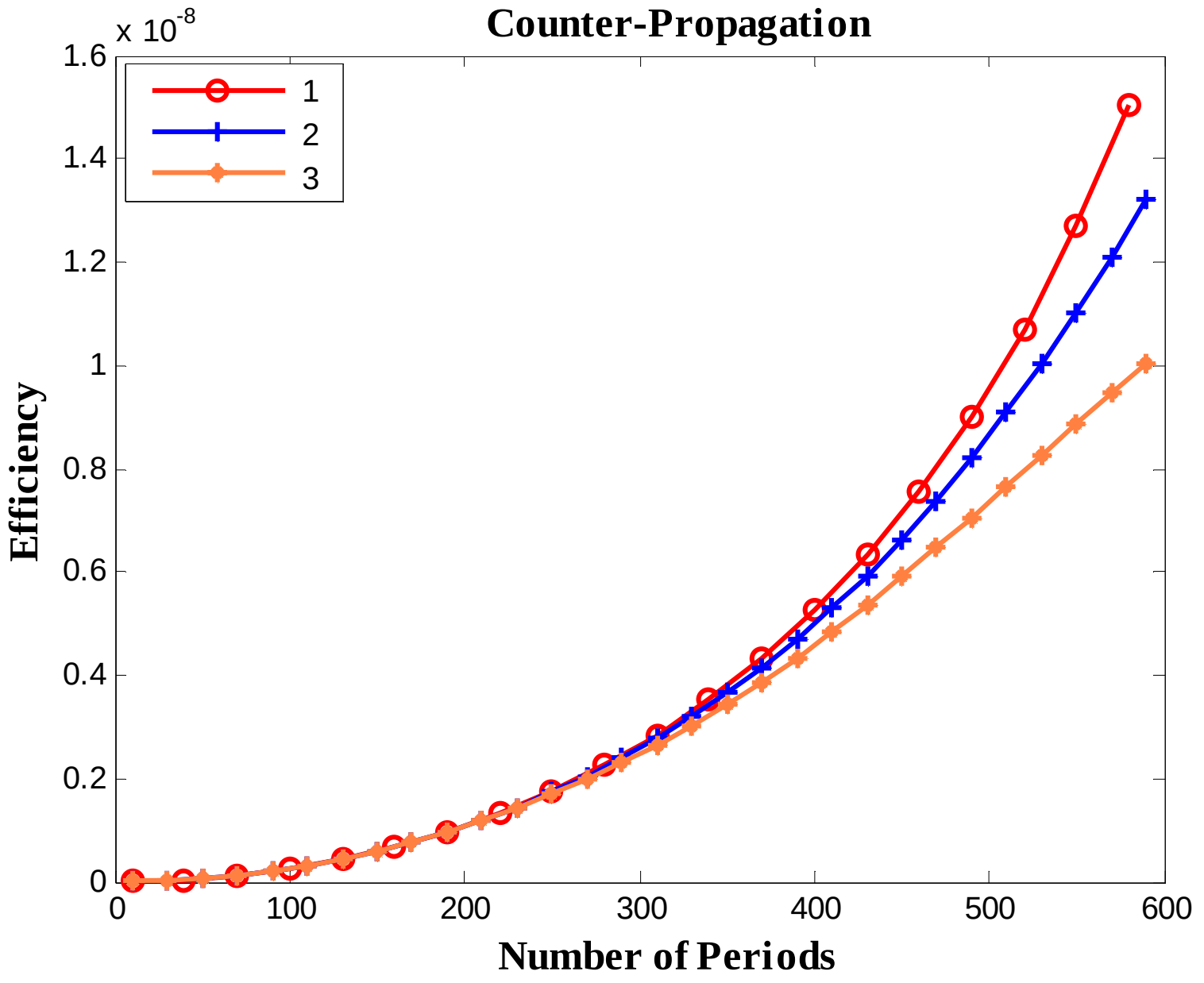

Fig. (8): Efficiency vs. number of periods for different values of $\delta$

1) $\delta = 0(m^{-1})$, 2) $\delta = 0.001k_0(m^{-1})$, 3) $\delta = 0.01k_0(m^{-1})$, $n_0 = 3.45, \Delta k = 0(m^{-1}), a_1 = 0.001, a_2 = -10^{-15}$

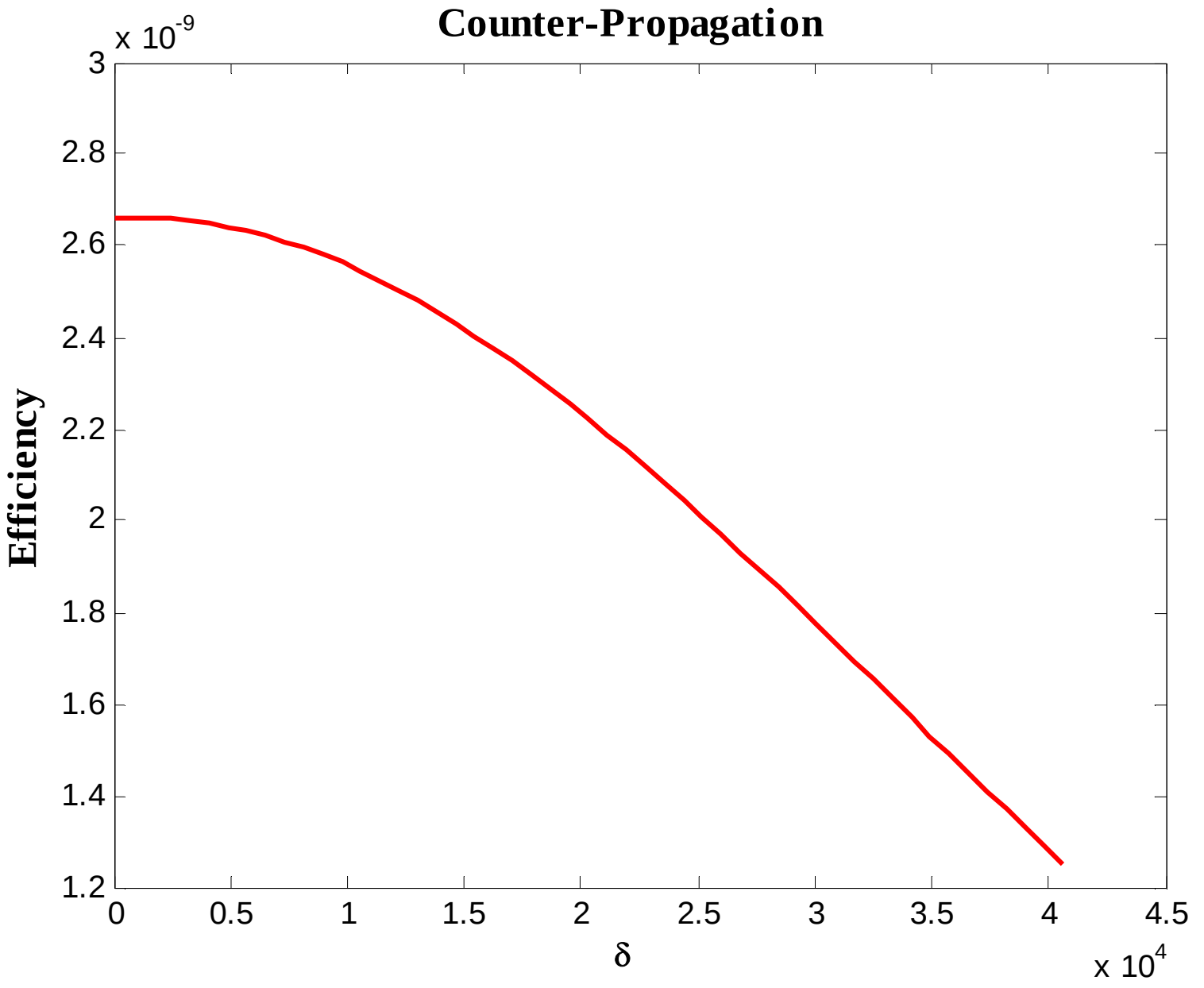

Fig. (9): Efficiency vs. phase mismatch between grating and average applied wave vectors

$n_0 = 3.45, \Delta k = 0(m^{-1}), a_1 = 0.001, a_2 = -10^{-15}, N = 300$

Fig (10) and (11) illustrate the effect of nonlinear refractive index's coefficient in different number of layers, on the counter-propagation conversion efficiency.

Conversion efficiency decreases by decreasing of the nonlinear refractive index's coefficient ($a_2$). FWM phenomenon is strongly affected by the nonlinear refractive index's coefficient and this effect has been shown in our numerical simulations.

Also, as we mentioned before, since the conversion efficiency is nonlinearly dependant on the number of periods, it is decreased when we choose more layers than 600 layers, so it is necessary to keep in mind this subject in design of a photonic crystal to avoid decreasing of the conversion efficiency.

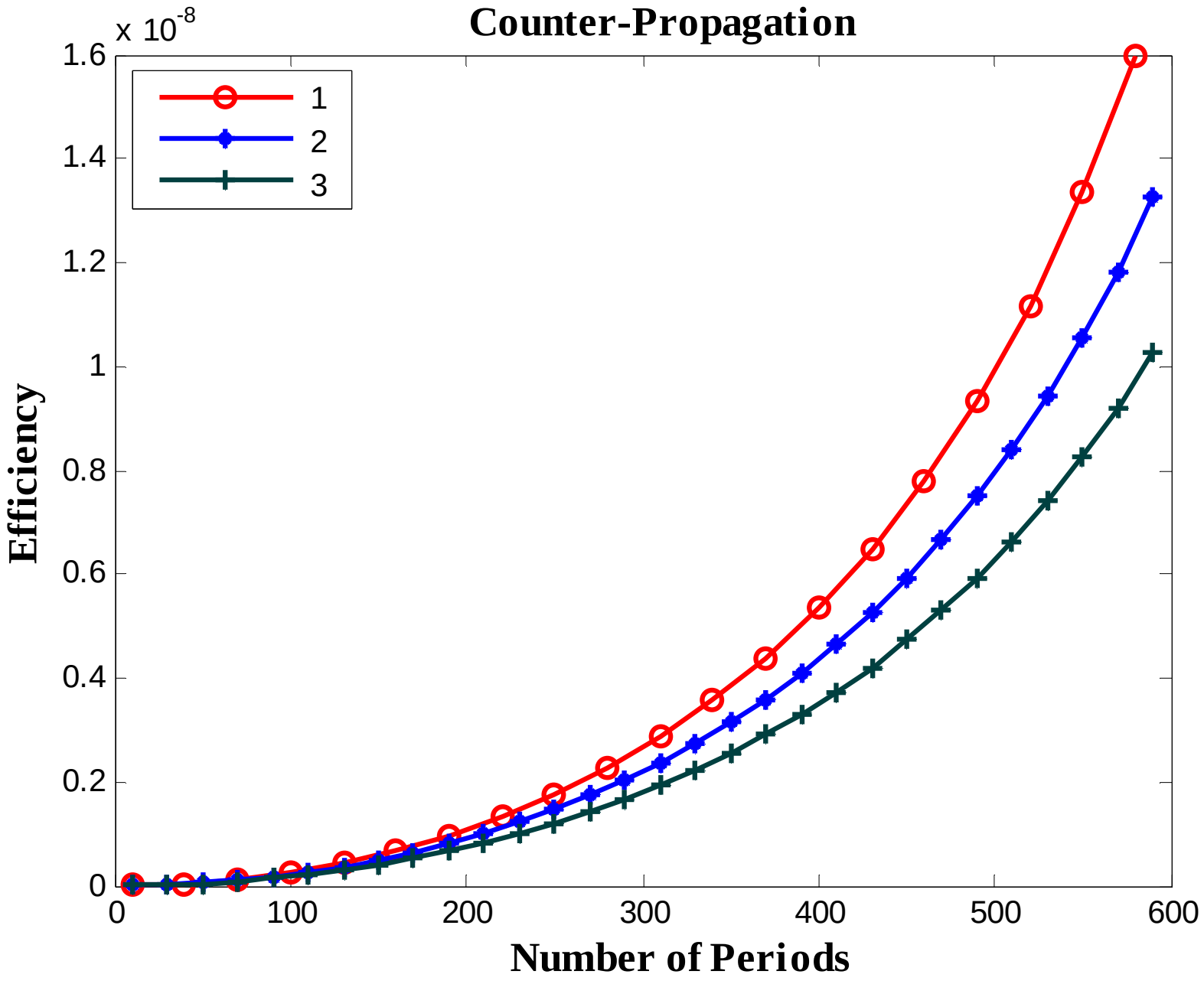

Fig. (10): Efficiency vs. number of periods for different values of $a_2$

1) $a_2 = -10^{-15}$, 2) $a_2 = -2\times10^{-15}$, 3) $a_2 = -3\times10^{-15}$, $n_0 = 3.45$, $\Delta k = 0(m^{-1})$, $a_1 = 0.001$, $\delta = 0(m^{-1})$

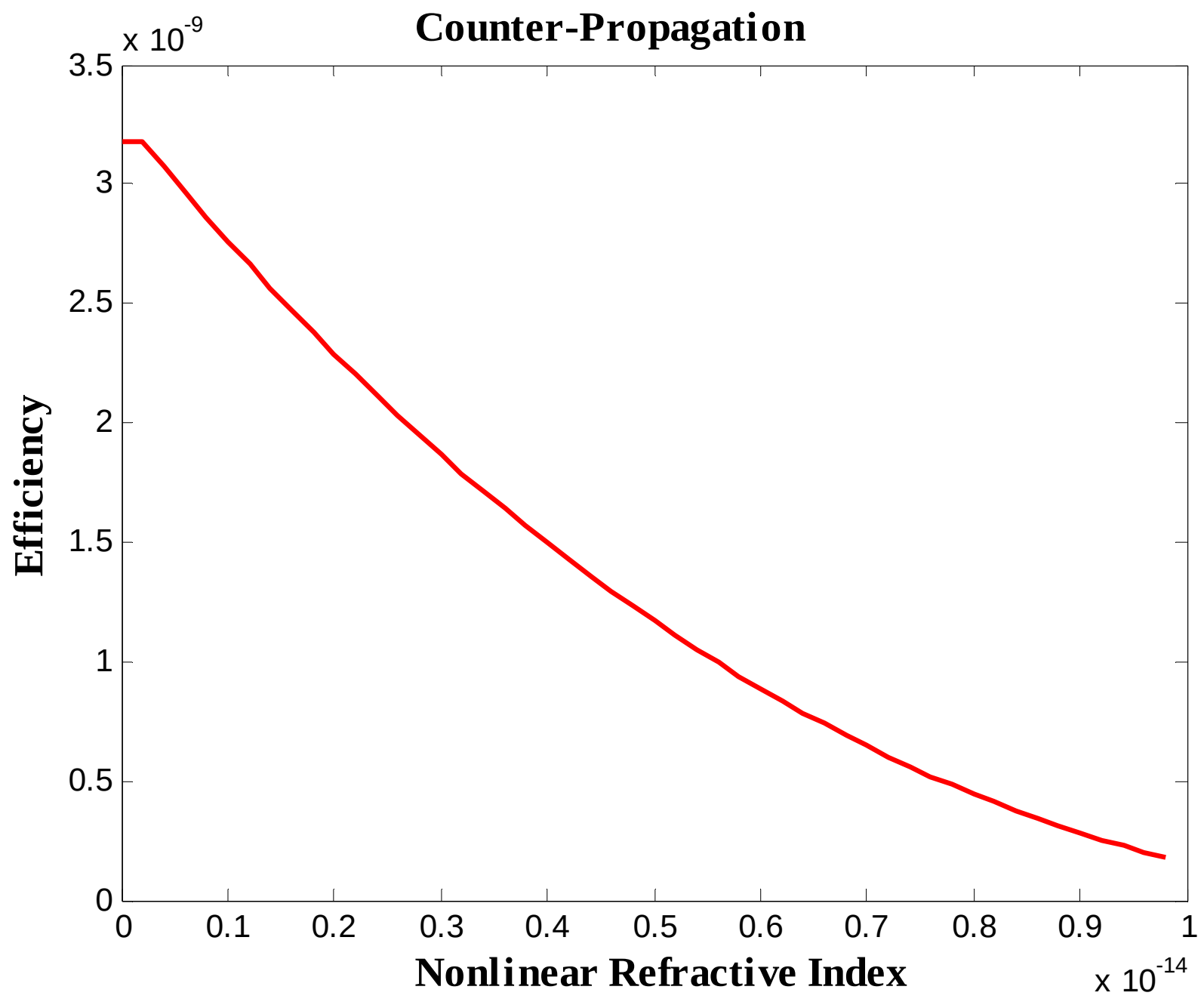

Fig. (11): Efficiency vs. absolute value of nonlinear refractive index's coefficient $n_0 = 3.45, \Delta k = 0(m^{-1}), a_1 = 0.001, \delta = 0(m^{-1}), N = 300$

**Summary and Conclusion-** The coupled mode equations for light propagation through one-dimensional nonlinear photonic crystals using FWM process have been derived and steady state numerical results have been discussed. We have considered the Bragg as a lossless and inhomogeneous medium. The linear and nonlinear refractive indexes are approximated with the first Fourier harmonic term.

It has been shown that although have considered that all the fields deplete via propagating in 1D photonic crystal and we have used a medium with $\chi^{(3)}$ nonlinearity, an enhancement of at least 100 times in both co-propagating and counter-propagating efficiencies have been observed compared to the previous works that use undepleted pump fields and $\chi^{(2)}$ medium which is larger than $\chi^{(3)}$ [15]

As to our knowledge this is the first time that NCMEs are derived via 1D nonlinear photonic crystal using the following cases:

1. We have used continuous wave frequency not short pulse case.
2. All fields (pump, signal and idler fields) are affected by Bragg grating, so both SPM and XPM phenomena are observed in Eqs.(9)-(23).
3. Pump depletion have been considered.
4. Both linear and nonlinear refractive of indexes are periodic.
5. In simulation processes both backward and forward components of the fields have been taken into account and none of them have been neglected.